%% file: main.tex
\documentclass[letterpaper, 10 pt, conference]{ieeeconf}
\IEEEoverridecommandlockouts 
\input{preamble}

\title{\LARGE \bf Efficient Reachability Analysis of Closed-Loop Systems \\with Neural Network Controllers}

\author{Michael Everett, Golnaz Habibi, Jonathan P. How
\thanks{The authors are with Aerospace Controls Laboratory at Massachusetts Institute of Technology, {\tt\small{\{mfe, ghabibi, jhow\}@mit.edu}}.}%
\thanks{This work was supported by Ford Motor Company.}%
\thanks{\textbf{Code:} \url{https://github.com/mit-acl/nn_robustness_analysis}}%
}

\begin{document}

\maketitle
\thispagestyle{empty}
\pagestyle{empty}

\begin{abstract}

Neural Networks (NNs) can provide major empirical performance improvements for robotic systems, but they also introduce challenges in formally analyzing those systems' safety properties.
In particular, this work focuses on estimating the forward reachable set of closed-loop systems with NN controllers.
Recent work provides bounds on these reachable sets, yet the computationally efficient approaches provide overly conservative bounds (thus cannot be used to verify useful properties), whereas tighter methods are too intensive for online computation.
This work bridges the gap by formulating a convex optimization problem for reachability analysis for closed-loop systems with NN controllers.
While the solutions are less tight than prior semidefinite program-based methods, they are substantially faster to compute, and some of the available computation time can be used to refine the bounds through input set partitioning, which more than overcomes the tightness gap.
The proposed framework further considers systems with measurement and process noise, thus being applicable to realistic systems with uncertainty.
Finally, numerical comparisons show that our approach based on linear programming and partitioning can give $10\times$ reduction in conservatism in $\frac{1}{2}$ of the computation time compared to the state-of-the-art, and the ability to handle various sources of uncertainty is highlighted on a quadrotor model.

\end{abstract}

\input{intro}
\input{related_work}
\input{background}
\input{approach}
\input{results}
\input{conclusion}

\bibliographystyle{ieeetr} 
\bibliography{main}

\end{document}

%% file: preamble.tex
\usepackage[utf8]{inputenc}
\usepackage{amsmath}
\usepackage{graphicx}
\usepackage{bm}
\usepackage{dsfont}
\usepackage{eufrak}
\usepackage{subcaption}
\DeclareCaptionLabelSeparator{periodspace}{.\quad}
\usepackage[dvipsnames]{xcolor}
\usepackage[sort,compress]{cite}
\usepackage{amsfonts}
\usepackage{url}
\usepackage[pdftex, pdfstartview={FitV}, pdfpagelayout={TwoColumnLeft},bookmarksopen=true,plainpages = false, colorlinks=true, linkcolor=black, citecolor = black, urlcolor = black,filecolor=black , pagebackref=false,hypertexnames=false, plainpages=false, pdfpagelabels ]{hyperref}
\usepackage{balance}
\usepackage[capitalize]{cleveref}

\newtheorem{theorem}{Theorem}[section]

\newtheorem{lemma}[theorem]{Lemma}

\crefname{section}{Section}{Sections}
\crefname{theorem}{Theorem}{Theorems}
\crefname{lemma}{Lemma}{Lemmas}
\crefname{table}{Table}{Tables}
\crefformat{equation}{(#2#1#3)}

\newcommand{\Aoutj}{\mathbf{A}_{\mathfrak{j},:}^{\textrm{out}}}
\newcommand{\xinX}{\mathbf{x}_t \in \mathcal{X}_\text{in}}
\newcommand{\piUCL}{\pi^{U_{CL}}_{:,\mathfrak{j}}}
\newcommand{\piLCL}{\pi^{L_{CL}}_{:,\mathfrak{j}}}
\newcommand{\Nuj}{\mathbf{n}^U_{\mathfrak{j}}}
\newcommand{\Muj}{\mathbf{M}^U_{\mathfrak{j},:}}
\newcommand{\Nlj}{\mathbf{n}^L_{\mathfrak{j}}}
\newcommand{\Mlj}{\mathbf{M}^L_{\mathfrak{j},:}}
\newcommand{\bpeps}{\mathcal{B}_p(\mathbf{x}_0,\epsilon)}
\newcommand{\Ainbin}{\mathbf{A}^{in}\mathbf{x}_t\leq \mathbf{b}^{in}}

\newcommand{\Aout}{\mathbf{A}^\text{out}}
\newcommand{\Ain}{\mathbf{A}^\text{in}}

\newcommand{\x}{\mathbf{x}}

\newcommand{\Au}[1]{\mathbf{\Lambda}^{#1}}
\newcommand{\Al}[1]{\mathbf{\Omega}^{#1}}

\newcommand{\upbias}[1]{\mathbf{\Delta}^{#1}}
\newcommand{\lwbias}[1]{\mathbf{\Theta}^{#1}}

\newcommand{\z}{\mathbf{z}}
\newcommand{\y}{\mathbf{y}}
\newcommand{\bias}[1]{\mathbf{b}^{#1}}

\newcommand{\R}{\mathds{R}}

\usepackage{accents}
\newcommand{\ubar}[1]{\underaccent{\bar}{#1}}

\DeclareMathOperator*{\argmax}{\arg\!\max}

\usepackage[addedmarkup=bf]{changes}
\definechangesauthor[name={MFE}, color={blue}]{MFE}
\definechangesauthor[name={GH}, color={ForestGreen}]{GH}
\definechangesauthor[name={JH}, color={red}]{JH}

%% file: intro.tex

\section{Introduction}\label{sec:intro}

\begin{figure}[t]
	\includegraphics[page=2,width=\linewidth,trim=250 150 240 70,clip]{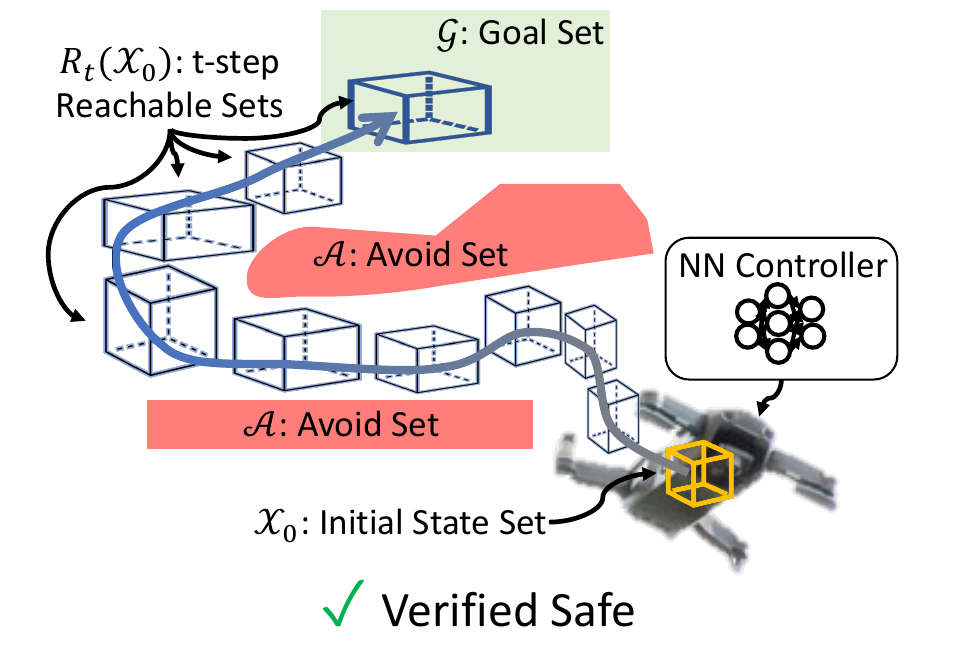}
	\caption{Forward Reachability Analysis. The objective is to compute the blue sets $\mathcal{R}_t(\mathcal{X}_0)$, to ensure a system starting from $\mathcal{X}_0$ (yellow) ends in $\mathcal{G}$ (green) and avoids $\mathcal{A}_0,\mathcal{A}_1$ (red). This is especially challenging for systems with NN control policies.}
	\label{fig:problem_cartoon}
	\vspace{-0.2in}
\end{figure}

Neural Networks (NNs) are pervasive in robotics due to their ability to express highly general input-output relationships for perception, planning, and control tasks.
However, before deploying NNs on safety-critical systems, there must be techniques to guarantee that the closed-loop behavior of systems with NNs will meet desired specifications.
The goal of this paper is to develop a framework for guaranteeing that systems with NN controllers will reach their goal states while avoiding undesirable regions of the state space, as in~\cref{fig:problem_cartoon}.

Despite the importance of analyzing closed-loop behavior, much of the recent work on formal NN analysis has focused on NNs in isolation (e.g., for image classification)~\cite{Ehlers_2017, Katz_2017, Huang_2017b,Lomuscio_2017,Tjeng_2019,Gehr_2018}, with an emphasis on efficiently relaxing NN nonlinearities~\cite{gowal2018effectiveness,Weng_2018,singh2018fast,zhang2018efficient,Wong_2018,Raghunathan_2018,fazlyab2019safety}.
On the other hand, closed-loop system reachability has been studied for decades, but traditional approaches, such as Hamilton-Jacobi methods~\cite{tomlin2000game,bansal2017hamilton}, do not consider NNs in the loop.

A handful of recent works~\cite{dutta2019reachability,huang2019reachnn,fan2020reachnn,ivanov2019verisig,xiang2020reachable,hu2020reach} propose methods that compute forward reachable sets of closed-loop systems with NN controllers.
A key challenge is in maintaining computational efficiency while still providing tight bounds on the reachable sets.
Moreover, the literature typically assumes perfect knowledge of system dynamics, with no stochasticity.

To address the primary challenge of computational efficiency, we re-formulate the semi-definite program (SDP) from~\cite{hu2020reach} as a linear program (LP) and leverage tools from~\cite{zhang2018efficient}.
While this relaxation provides substantial improvement in computational efficiency, it also introduces some conservatism.
Thus, the proposed algorithm trades off some computational efficiency for bound tightness by partitioning the input set, as motivated by~\cite{xiang2018reachable}.
Finally, the proposed framework considers measurement and process noise throughout the formulation, thus being more amenable to applications on real, uncertain closed-loop systems.

This work's contributions include:
(i) a convex optimization formulation of reachability analysis for closed-loop systems with NN controllers, providing a computationally efficient method for verifying safety properties,
(ii) the use of input set partitioning techniques to provide tight bounds on the reachable sets despite large initial state sets,
(iii) the consideration of measurement and process noise, which improves the applicability to real systems with uncertainty,
and (iv) numerical comparisons with~\cite{hu2020reach} showing 10$\times$ tighter accuracy in $\frac{1}{2}$ the computation time via LP and partitioning, and an application on noisy quadrotor dynamics.

%% file: related_work.tex

\section{Related Work}\label{sec:related_work}
Related work on reachability analysis can be categorized into works on NNs in isolation, closed-loop systems without NNs, and closed-loop systems with NNs.
For instance, machine learning literature includes many methods to verify properties of NNs, often motivated by defending against adversarial examples~\cite{Szegedy_2014}.
These methods broadly range from exact~\cite{katz2017reluplex} to tight~\cite{fazlyab2019safety} to efficient~\cite{zhang2018efficient} to fast~\cite{gowal2018effectiveness}.
Although these tools are not designed for closed-loop systems, the  NN relaxations from \cite{zhang2018efficient} provide a key foundation here.

For closed-loop systems, reachability analysis is a standard component of safety verification.
Modern methods include Hamilton-Jacobi Reachability methods~\cite{tomlin2000game,bansal2017hamilton}, SpaceEx~\cite{frehse2011spaceex}, Flow*~\cite{chen2013flow}, CORA~\cite{althoff2015introduction}, and C2E2~\cite{duggirala2015c2e2,fan2016automatic}, but these do not account for NN control policies.
Orthogonal approaches that do not explicitly estimate the system's forward reachable set, but provide other notions of safety, include Lyapunov function search~\cite{papachristodoulou2002construction} and control barrier functions (CBFs)~\cite{ames2016control}.

Recent reachability analysis approaches that do account for NN control policies face a tradeoff between computation time and conservatism.
\cite{dutta2019reachability,huang2019reachnn,fan2020reachnn} use polynomial approximations of NNs to make the analysis tractable.
Most works consider NNs with ReLU approximations, whereas \cite{ivanov2019verisig} considers sigmoidal activations.
\cite{xiang2020reachable,yang2019efficient} introduce conservatism by assuming the NN controller could output its extreme values at every state.
Most recently, \cite{hu2020reach} formulated the problem as a SDP, called Reach-SDP.
This work builds on both~\cite{xiang2020reachable,hu2020reach} and makes the latter more scalable by re-formulating the SDP as a linear program, introduces sources of uncertainty in the closed-loop dynamics, and shows further improvements by partitioning the input set.

%% file: background.tex

\section{Preliminaries}

\subsection{Closed-Loop System Dynamics}

Consider a discrete-time linear time-varying system,
\begin{align}
\begin{split}
    \mathbf{x}_{t+1} &= A_t \mathbf{x}_{t} + B_t \mathbf{u}_t + \mathbf{c}_t + \bm{\omega}_t \label{eqn:ltv_dynamics} \\
    \mathbf{y}_{t} &= C_t^T\mathbf{x}_t + \bm{\nu}_t,
\end{split}
\end{align}
where $\mathbf{x}_t\in\R^{n_x},\, \mathbf{u}_t\in\R^{n_u},\, \mathbf{y}_t\in\R^{n_y}$ are state, control, and output vectors, $A_t,\, B_t,\, C_t$ are known system matrices, $\mathbf{c}_t\in\R^{n_x}$ is a known exogenous input, and $\bm{\omega}_t\sim\Omega$ and $\bm{\nu}_t\sim N$ are process and measurement noises sampled at each timestep from unknown distributions with known, finite support (i.e., $\bm{\omega}_t\in [\ubar{\bm{\omega}}_t,\, \bar{\bm{\omega}}_t ], \bm{\nu}_t\in [\ubar{\bm{\nu}}_t,\, \bar{\bm{\nu}}_t ]$ element-wise).

We assume an output-feedback controller $\mathbf{u}_t=\pi(\mathbf{y}_t)$ parameterized by an $m$-layer feed-forward NN, optionally subject to control constraints, $\mathbf{u}_t\in\mathcal{U}_t$.
We denote the closed-loop system with dynamics~\cref{eqn:ltv_dynamics} and control policy $\pi$ as
\begin{equation}
    \mathbf{x}_{t+1} = f(\mathbf{x}_{t}; \pi). \label{eqn:closed_loop_dynamics}
\end{equation}

\subsection{Reachable Sets}

For the closed-loop system~\cref{eqn:closed_loop_dynamics}, we denote $\mathcal{R}_t(\mathcal{X}_0)$ the forward reachable set at time $t$ from a given set of initial conditions $\mathcal{X}_0\subseteq\R^{n_x}$, which is defined by the recursion
\begin{equation}
    \mathcal{R}_{t+1}(\mathcal{X}_0) = f(\mathcal{R}_t(\mathcal{X}_0); \pi), \qquad \mathcal{R}_0(\mathcal{X}_0)=\mathcal{X}_0.\label{eqn:reachable_sets}
\end{equation}

\subsection{Finite-Time Reach-Avoid Verification Problem}

The finite-time reach-avoid properties verification is defined as follows: Given a goal set $\mathcal{G}\subseteq\R^{n_x}$, a sequence of avoid sets $\mathcal{A}_t\subseteq\R^{n_x}$, and a sequence of reachable set estimates $\mathcal{R}_t\subseteq\R^{n_x}$, determining that every state in the final estimated reachable set will be in the goal set and any state in the estimated reachable sets will not enter an avoid set requires computing set intersections, $\texttt{VERIFIED}(\mathcal{G},\mathcal{A}_{0:N},\mathcal{R}_{0:N})\equiv{\mathcal{R}_N\subseteq\mathcal{G}}\ \mathrm{\&}\ {\mathcal{R}_t\cap\mathcal{A}_t=\emptyset}, \forall t\in\{0,\ldots,N\}$.

In the case of our nonlinear closed-loop system~\cref{eqn:closed_loop_dynamics}, where computing the reachable sets exactly is computationally intractable, we can instead compute outer-approximations of the reachable sets, $\bar{\mathcal{R}}(\mathcal{X}_0)\supseteq\mathcal{R}_t(\mathcal{X}_0)$.
This is useful if the finite-time reach-avoid properties of the system as described by outer-approximations of the reachable sets are verified, because that implies the finite-time reach-avoid properties of the \textit{exact} closed loop system are verified as well.
Tight outer-approximations of the reachable sets are desirable, as they enable verification of tight goal and avoid set specifications, and they reduce the chances of verification being unsuccessful even if the exact system meets the specifications.

\subsection{Control Policy Neural Network Structure}

Using notation from~\cite{zhang2018efficient}, for the $m$-layer neural network used in the control policy, the number of neurons in each layer is $n_k \forall k\in [m]$, where $[i]$ denotes the set $\{1,2,\ldots,i\}$.
Let the $k$-th layer weight matrix be $\mathbf{W}^{(k)}\in\R^{n_k\times n_{k-1}}$ and bias vector be $\mathbf{b}^{(k)}\in\R^{n_k}$, and let $\Phi_k: \R^{n_x}\to\R^{n_k}$ be the operator mapping from network input (measured output vector $\mathbf{y}_t$) to layer $k$.
We have $\Phi_k(\mathbf{y}_t)=\sigma(\mathbf{W}^{(k)}\Phi_{k-1}(\mathbf{y}_t)+\mathbf{b}^{(k)}),\forall k\in [m-1]$, where $\sigma(\cdot)$ is the coordinate-wise activation function.
The framework applies to general activations, including ReLU, $\sigma(\mathbf{z})=\mathrm{max}(0,\mathbf{z})$.
The network input $\Phi_0(\mathbf{y}_t)=\mathbf{y}_t$ produces the unclipped control input,
\begin{equation}
    \mathbf{u}_t=\pi(\mathbf{y}_t)=\Phi_m(\mathbf{y}_t)=\mathbf{W}^{(m)}\Phi_{m-1}(\mathbf{y}_t)+\mathbf{b}^{(m)}.
\end{equation}

\subsection{Neural Network Robustness Verification}

A key step in quickly computing reachable sets of the closed-loop system~\cref{eqn:closed_loop_dynamics} with a NN control policy is to relax nonlinear constraints induced by the NN's nonlinear activation functions.
Within a known range of a neuron's input, a nonlinear activation can be linearly bounded above/below.

\begin{theorem}[From~\cite{zhang2018efficient}, Convex Relaxation of NN]\label{thm:crown_particular_x}
Given an $m$-layer neural network control policy $\pi:\R^{n_y}\to\R^{n_u}$, there exist two explicit functions $\pi_j^L: \R^{n_y}\to\R^{n_u}$ and $\pi_j^U: \R^{n_y}\to\R^{n_u}$ such that $\forall j\in [n_m], \forall \mathbf{y}\in\mathcal{B}_p(\mathbf{y}_0, \epsilon)$, the inequality $\pi_j^L(\mathbf{y})\leq \pi_j(\mathbf{y})\leq \pi_j^U(\mathbf{y})$ holds true, where
\begin{align}
\label{eq:f_j_UL}
    \pi_{j}^{U}(\y) &= \Au{(0)}_{j,:} \y + \sum_{k=1}^{m}\Au{(k)}_{j,:}(\bias{(k)}+\upbias{(k)}_{:,j}) \\
    \pi_{j}^{L}(\y) &= \Al{(0)}_{j,:} \y + \sum_{k=1}^{m}\Al{(k)}_{j,:}(\bias{(k)}+\lwbias{(k)}_{:,j}),
\end{align}
where $\Au{}, \Al{}, \upbias{}, \lwbias{}$ are defined recursively using NN weights and activations (e.g., ReLU, tanh), as detailed in~\cite{zhang2018efficient}.
\end{theorem}

In a closed-loop system, \cref{thm:crown_particular_x} bounds the control output for a \textit{particular} measurement $\mathbf{y}$.
Moreover, if all that is known is $\mathbf{y}\in\mathcal{B}_p(\mathbf{y}_0, \epsilon)$, \cref{thm:crown_particular_x} provides affine relationships between $\mathbf{y}$ and $\mathbf{u}$ (i.e., bounds valid within the known set of possible $\y$).
These relationships enable efficient calculation of NN output bounds, using Corollary 3.3 of~\cite{zhang2018efficient}.

We could leverage~\cite{zhang2018efficient} to compute reachable sets by first bounding the possible controls, then bounding the next state set by applying the extreme controls from each state.
This is roughly the approach in~\cite{xiang2020reachable,yang2019efficient}, for example.
However, this introduces excessive conservatism, because both extremes of control would not be applied at every state (barring pathological examples).
To produce tight bounds on the reachable sets, we leverage the relationship between measured output and control in~\cref{sec:approach}.

%% file: approach.tex

\section{Approach}\label{sec:approach}

\begin{figure*}[t]
    \centering
    \includegraphics[page=3,width=0.8\linewidth,trim=0 170 0 110,clip]{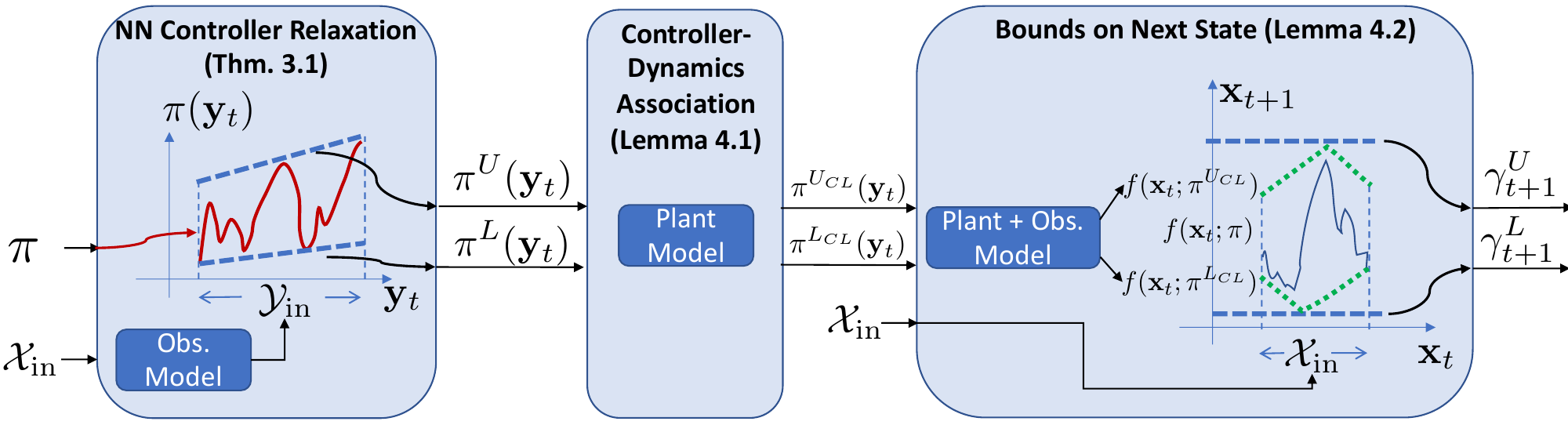}
    \caption{Approach Overview for simple 1D system. \cref{thm:crown_particular_x} relaxes the NN to give affine relationships between observation $\y_t$ and control: $\pi^U, \pi^L$. \cref{thm:particular_xt} uses the system dynamics to associate $\pi^U, \pi^L$ with the next state set. \cref{thm:bounds_on_xt1} optimizes the closed-loop dynamics over all states $\xinX$ to compute bounds on the next state, $\gamma^U_{t+1}, \gamma^L_{t+1}$.}
    \label{fig:approach_flow}
    \vspace{-0.2in}
\end{figure*}

Recall that our goal is to find the set of all possible next states, $\mathbf{x}_{t+1}\in\mathcal{X}_\text{out}$, given that the current state lies within a known set, $\mathbf{x}_t \in \mathcal{X}_\text{in}$.
This will allow us to compute reachable sets recursively starting from an initial set $\mathcal{X}_\text{in}=\mathcal{X}_0$.

The approach follows the architecture in~\cref{fig:approach_flow}.
After first relaxing the NN controller using~\cref{thm:crown_particular_x}, we then associate linearized extreme controllers with extreme next states in~\cref{sec:approach:particular_xt}.
Then, using the linearized extreme controller, we optimize over all states in the input set to find extreme next states in~\cref{sec:approach:bounds_xt1}.
We extend the formulation to handle control limits in~\cref{sec:approach:control_limits}, then describe how to convert the solutions of the optimization problems into reachable set descriptions in~\cref{sec:approach:reachable_sets}.

\subsection{Assumptions}
This work assumes that $\mathcal{X}_\text{in}$ is described by either:
\begin{itemize}
	\item an $\ell_p$ ball for some norm $p\in[1,\infty]$ and radius $\epsilon$, s.t. $\mathcal{X}_\text{in}=\bpeps$; or
	\item a polytope, for some $\Ain\in\R^{m_\text{in}\times n_x}$, $\mathbf{b}^\text{in}\in\R^{m_\text{in}}$, s.t. $\mathcal{X}_\text{in}=\{\mathbf{x}_t \lvert \Ainbin\}$,
\end{itemize}
and shows how to compute $\mathcal{X}_\text{out}$ as described by either:
\begin{itemize}
	\item an $\ell_{\infty}$ ball with radius $\epsilon$, s.t. $\mathcal{X}_\text{out}=\mathcal{B}_{\infty}(\mathbf{x}_0,\epsilon)$; or
	\item a polytope for a specified $\Aout\in\mathds{R}^{m_\text{out}\times n_x}$, meaning we will compute $\mathbf{b}^\text{out}\in\mathds{R}^{m_{out}}$ s.t. $\mathcal{X}_\text{out}=\{\mathbf{x}_t \in \mathds{R}^{n_x} \lvert \Aout\mathbf{x}_t\leq \mathbf{b}^\text{out}\}$.
\end{itemize}

We assume that either $\Aout$ is provided (in the case of polytope output bounds), or that $\Aout=\mathbf{I}_{n_x}$ (in the case of $\ell_{\infty}$ output bounds).
Note that we use $\mathsf{j}$ to index the state vectors, $\mathfrak{j}$ to index polytope facets, and $j$ to index the control vectors.
\cref{sec:approach:particular_xt,sec:approach:bounds_xt1} assume that $\mathcal{U}_t=\mathds{R}^{n_u}$ (no control input constraints) for cleaner notation; this assumption is relaxed in~\cref{sec:approach:control_limits}.

\subsection[Bounds on Next State from a particular state]{Bounds on $\mathbf{x}_{t+1}$ from a particular $\mathbf{x}_t$}\label{sec:approach:particular_xt}


\begin{lemma}\label[lemma]{thm:particular_xt}
Given an $m$-layer NN control policy $\pi:\R^{n_y}\to\R^{n_u}$, closed-loop dynamics $f: \R^{n_x} \times \Pi \to \R^{n_x}$ as in~\cref{eqn:ltv_dynamics,eqn:closed_loop_dynamics}, and specification matrix $\Aout\in\R^{n_\text{out}\times n_x}$, for each $\mathfrak{j}\in [n_\text{out}]$, there exist two explicit functions $\piLCL: \R^{n_y}\to\R^{n_u}$ and $\piUCL: \R^{n_y}\to\R^{n_u}$ such that $\forall j\in [n_m], \forall \x_t\in\mathcal{B}_p(\x_{t,0}, \epsilon)$ and $\forall \y_t\in\mathcal{B}_\infty(C_t^T\x_t + \frac{\bar{\bm{\nu}}_t+\ubar{\bm{\nu}}_t}{2}, \frac{\bar{\bm{\nu}}_t-\ubar{\bm{\nu}}_t}{2})$, the inequality $\Aoutj f(\x_t, \piLCL) \leq \Aoutj f(\x_t, \pi) \leq \Aoutj f(\x_t, \piUCL)$ holds true, where
\begin{align}
    \piUCL(\y_t) &= \mathbf{\Upsilon}^{(0)}_{\mathfrak{j},:,:}\y_t+ \z^U \label{eqn:piucl}\\
    \piLCL(\y_t) &= \mathbf{\Xi}^{(0)}_{\mathfrak{j},:,:}\y_t+ \z^L \label{eqn:pilcl},
\end{align}
letting
\begin{align}
    \z^U & = \sum_{k=1}^{m} \left[\mathbf{\Upsilon}^{(k)}_{\mathfrak{j},:,:} \mathbf{b}^{(k)}+\mathds{1}_{n_u}\left(\left(\mathbf{\Upsilon}^{(k)}_{\mathfrak{j},:,:}\right)^T\odot\mathbf{\Psi}^{(k)}\right)\right] \\
    \z^L & = \sum_{k=1}^{m} \left[\mathbf{\Xi}^{(k)}_{\mathfrak{j},:,:} \mathbf{b}^{(k)}+\mathds{1}_{n_u}\left(\left(\mathbf{\Xi}^{(k)}_{\mathfrak{j},:,:}\right)^T\odot\mathbf{\Gamma}^{(k)}\right)\right]
\end{align}
and $\forall k\in[m]$, $\mathbf{\Upsilon}^{(k)}\in\R^{m_\text{out}\times n_u \times n_u}, \mathbf{\Psi}^{(k)}\in\R^{m_\text{out}\times n_k \times n_u}$,
\begin{align}
    \mathbf{\Upsilon}_{\mathfrak{j},:,:}^{(k)} &= \bar{\mathbf{J}}_{\mathfrak{j},:,:}^{(k)} \mathbf{\Lambda}^{(k)} + \ubar{\mathbf{J}}_{\mathfrak{j},:,:}^{(k)} \mathbf{\Omega}^{(k)}\\
    \mathbf{\Psi}_{\mathfrak{j},:,:}^{(k)} &= \mathbf{\Delta}^{(k)} \bar{\mathbf{J}}_{\mathfrak{j},:,:}^{(k)} + \mathbf{\Theta}^{(k)} \ubar{\mathbf{J}}_{\mathfrak{j},:,:}^{(k)}\\
    \mathbf{\Xi}_{\mathfrak{j},:,:}^{(k)} &= \bar{\mathbf{J}}_{\mathfrak{j},:,:}^{(k)} \mathbf{\Omega}^{(k)} + \ubar{\mathbf{J}}_{\mathfrak{j},:,:}^{(k)} \mathbf{\Lambda}^{(k)}\\
    \mathbf{\Gamma}_{\mathfrak{j},:,:}^{(k)} &= \mathbf{\Theta}^{(k)} \bar{\mathbf{J}}_{\mathfrak{j},:,:}^{(k)} + \mathbf{\Delta}^{(k)} \ubar{\mathbf{J}}_{\mathfrak{j},:,:}^{(k)}.
\end{align}
using selector matrices $\bar{\mathbf{J}}^{(k)},\ubar{\mathbf{J}}^{(k)}\in\{0,1\}^{n_u\times n_k\times n_u}$,
\begin{align}
    \bar{\mathbf{J}}_{\mathfrak{j},j,:}^{(k)} &=
    \begin{cases}
        \mathbf{e}_{j}^{T}, & \textrm{if } \Aoutj B_{t,:,\mathfrak{j}} \geq 0 \\
        \mathbf{0}^{T}, & \textrm{otherwise}
    \end{cases} \label{eqn:selector_above}\\
    \ubar{\mathbf{J}}_{\mathfrak{j},j,:}^{(k)} &=
    \begin{cases}
        \mathbf{0}^{T}, & \textrm{if } \Aoutj B_{t,:,\mathfrak{j}} \geq 0 \\
        \mathbf{e}_{j}^{T}, & \textrm{otherwise}
    \end{cases}, \label{eqn:selector_below}
\end{align}
and $\Au{}, \Al{}, \upbias{}, \lwbias{}$ are computed from~\cref{thm:crown_particular_x} with $\y_0=C_t^T(\x_{t,0}+\frac{\bar{\bm{\nu}}_t+\ubar{\bm{\nu}}_t}{2})$, and $\epsilon=\epsilon+\frac{\bar{\bm{\nu}}_t-\ubar{\bm{\nu}}_t}{2}$.

\begin{proof}
For any particular measurement $\y_t$, after relaxing the NN according to~\cref{thm:crown_particular_x}, let $\Pi(\y_t)=\{ \pi \lvert \pi_j^L(\y_t) \leq \pi_j(\y_t) \leq \pi_j^U(\y_t) \forall j\in[n_u] \}$ denote the set of possible effective control policies.
Denote the control policy $\piUCL\in\Pi(\y_t)$ as one that induces the least upper bound on the $\mathfrak{j}$-th facet of the next state polytope,
\begin{align}
    \Aoutj f(\mathbf{x}_{t}; \pi_{:,\mathfrak{j}}^{U_{CL}}) =& \max_{\pi \in \Pi(\y_t)} \Aoutj  f(\mathbf{x}_t; \pi) \notag \\
    & \hspace*{-1.1in} = \max_{\pi \in \Pi(\y_t)} \Aoutj \left[A_{t} \mathbf{x}_t +  
    B_{t} \pi(\y_t) + \mathbf{c}_{t} + \bm{\omega}_t\right] \nonumber\\
    &\hspace*{-1.1in} = \left[\max_{\pi \in \Pi(\y_t)} \Aoutj  B_{t} \pi(\y_t) \right] + 
    \Aoutj \left[A_{t} \mathbf{x}_t + \mathbf{c}_{t} + \bm{\omega}_t\right],
\end{align}
Thus for $\y_t$,
\begin{align}
    \pi_{:,\mathfrak{j}}^{U_{CL}} =& \argmax_{\pi \in \Pi(\y_t)} \Aoutj  B_{t} \pi(\y_t).
\end{align}
The resulting control input $\forall \mathfrak{j}\in[m_{t+1}],j\in[n_u]$ is,
\begin{align}
    \pi_{j,\mathfrak{j}}^{U_{CL}}(\y_t) =& \begin{cases}
        \pi_j^U(\y_t), & \mathrm{if}\ \Aoutj  B_{t,:,j} \geq 0 \\
        \pi_j^L(\y_t), & \mathrm{otherwise}
    \end{cases}
    \label{eqn:pi_upoly_elementwise}.
\end{align}
Writing \cref{eqn:pi_upoly_elementwise} in matrix form results in~\cref{eqn:piucl}.
The proof of the lower bound follows similarly.
\end{proof}
\end{lemma}

\subsection[Bounds on Next state from initial state set]{Bounds on $\x_{t+1}$ from any $\xinX$}\label{sec:approach:bounds_xt1}

Now that we can bound each facet of the next state polytope given a particular current state and observation, we can form bounds on the next state polytope facet given a \textit{set} of possible current states.
This is necessary to handle initial state set constraints and to compute ``$t>1$''-step reachable sets recursively as in~\cref{eqn:reachable_sets}.
We assume $\xinX$.

\begin{lemma}\label[lemma]{thm:bounds_on_xt1}
Given an $m$-layer NN control policy $\pi:\R^{n_y}\to\R^{n_u}$, closed-loop dynamics $f: \R^{n_x} \times \Pi \to \R^{n_x}$ as in~\cref{eqn:ltv_dynamics,eqn:closed_loop_dynamics}, and specification matrix $\Aout\in\R^{n_\text{out}\times n_x}$, for each $\mathfrak{j}\in[n_\text{out}]$, there exist two fixed values $\gamma_{t+1,\mathfrak{j}}^U$ and $\gamma_{t+1,\mathfrak{j}}^L$ such that $\forall \xinX$, the inequality $\gamma_{t+1,\mathfrak{j}}^L \leq \Aoutj f(\x_t; \pi) \leq \gamma_{t+1,\mathfrak{j}}^U$ holds true, where
\begin{align}
    \gamma_{t+1,\mathfrak{j}}^{U} &= \max_{\xinX} \Muj \mathbf{x}_t + \Nuj \label{eqn:global_upper_bnd_generic_optimization} \\
    \gamma_{t+1,\mathfrak{j}}^{L} &= \min_{\xinX} \Mlj \mathbf{x}_t + \Nlj\label{eqn:global_lower_bnd_generic_optimization},
\end{align}
with $\mathbf{M}^U\in\R^{n_x \times n_x}$, $\mathbf{n}^U\in\R^{n_x}$ defined as
\begin{align}
    \Muj &= \left(\Aoutj  \left(A_{t} + B_{t} \mathbf{\Upsilon}_{\mathsf{j},:,:}^{(0)} C_t^T \right) \right) \label{eqn:muj_defn}\\
    \Mlj &= \left(\Aoutj  \left(A_{t} + B_{t} \mathbf{\Psi}_{\mathsf{j},:,:}^{(0)} C_t^T \right) \right) \\
    \Nuj &= \Aoutj \left(B_{t}\left( \mathbf{\Upsilon}_{\mathsf{j},:,:}^{(0)}\left( \bar{\bar{\mathbf{J}}}_{\mathfrak{j},:,:}^{(0)}\bar{\bm{\nu}}_t + \ubar{\ubar{\mathbf{J}}}_{\mathfrak{j},:,:}^{(0)}\ubar{\bm{\nu}}_t \right) + \z^U\right) + \nonumber\right.\\ &\quad\quad\quad\quad\left.\mathbf{c}_{t} + \bar{\mathbf{J}}_{\mathfrak{j},:,:}^{(0)}\bar{\bm{\omega}}_t + \ubar{\mathbf{J}}_{\mathfrak{j},:,:}^{(0)}\ubar{\bm{\omega}}_t \right) \label{eqn:nuj_defn}\\
    \Nlj &= \Aoutj \left(B_{t}\left( \mathbf{\Psi}_{\mathsf{j},:,:}^{(0)}\left( \ubar{\ubar{\ubar{\mathbf{J}}}}_{\mathfrak{j},:,:}^{(0)}\bar{\bm{\nu}}_t + \bar{\bar{\bar{\mathbf{J}}}}_{\mathfrak{j},:,:}^{(0)}\ubar{\bm{\nu}}_t \right) + \z^U\right) + \nonumber\right.\\ &\quad\quad\quad\quad\left.\mathbf{c}_{t} + \ubar{\mathbf{J}}_{\mathfrak{j},:,:}^{(0)}\bar{\bm{\omega}}_t + \bar{\mathbf{J}}_{\mathfrak{j},:,:}^{(0)}\ubar{\bm{\omega}}_t \right),
\end{align}
and where $\{\bar{\bar{\mathbf{J}}}, \ubar{\ubar{\mathbf{J}}}\}$, $\{\bar{\bar{\bar{\mathbf{J}}}}, \ubar{\ubar{\ubar{\mathbf{J}}}}\}$ are defined as in~\cref{eqn:selector_above,eqn:selector_below}, but using $\Aoutj B_{t,:,\mathfrak{j}} \mathbf{\Upsilon}_{\mathsf{j},:,:}^{(0)} C_t^T$ and $\Aoutj B_{t,:,\mathfrak{j}} \mathbf{\Psi}_{\mathsf{j},:,:}^{(0)} C_t^T$, respectively,
with $\mathbf{\Upsilon}, \mathbf{\Psi}, \z^U, \z^L, \bar{\mathbf{J}}, \ubar{\mathbf{J}}$ computed from~\cref{thm:particular_xt}.

\begin{proof}
Bound the next state polytope's $\mathfrak{j}$-th facet above,
\begingroup
\allowdisplaybreaks
\begin{align}
    \Aoutj &\mathbf{x}_{t+1}=\Aoutj f(\mathbf{x}_{t};\pi) \\
    &\leq \Aoutj f(\mathbf{x}_{t}; \piUCL) \\
    &\leq \max_{\xinX}\Aoutj f(\mathbf{x}_{t}; \piUCL) := \gamma_{t+1,\mathfrak{j}}^U\\
    &= \max_{\xinX} \Aoutj  \left[A_{t} \mathbf{x}_t + B_{t} \piUCL(\y_t)+c_{t} + \bm{\omega}_t \right]\\
    &= \max_{\xinX} \Aoutj \left[A_{t} \mathbf{x}_t + B_{t} \left(\mathbf{\Upsilon}^{(0)}_{\mathfrak{j},:,:}\y_t+\z^U\right) + \nonumber\right.\\&\quad\quad\quad\quad\quad\quad\quad\left.c_{t} + \bm{\omega}_t \right] \label{eqn:global_bnd_sub_piUCL} \\ 
    &= \max_{\xinX} \Aoutj \left[A_{t} \mathbf{x}_t + B_{t} \left(\mathbf{\Upsilon}^{(0)}_{\mathfrak{j},:,:}\left(C_t^T\x_t+\bm{\nu}_t\right)+\z^U\right) \nonumber\right.\\&\quad\quad\quad\quad\quad\quad\quad\left.+ c_{t} + \bm{\omega}_t \right] \label{eqn:global_bnd_observation} \\
    &= \max_{\xinX} \left(\Aoutj \left(A_{t} + B_{t} \mathbf{\Upsilon}^{(0)}_{\mathfrak{j},:,:} C_t^T\right) \right) \x_t + \nonumber\\& \quad\quad\quad\Aoutj\left( B_{t} \left(\mathbf{\Upsilon}^{(0)}_{\mathfrak{j},:,:} \bm{\nu}_t + \z^U \right) + c_{t} + \bm{\omega}_t \right) \label{eqn:global_bnd_linear_in_x} \\
    &= \max_{\xinX} \left(\Aoutj \left(A_{t} + B_{t} \mathbf{\Upsilon}^{(0)}_{\mathfrak{j},:,:} C_t^T\right) \right) \x_t + \nonumber\\& \quad\quad\quad\Aoutj\left( B_{t} \left(\mathbf{\Upsilon}^{(0)}_{\mathfrak{j},:,:} \left(\bar{\bar{\mathbf{J}}}_{\mathfrak{j},:,:}^{(0)}\bar{\bm{\nu}}_t + \ubar{\ubar{\mathbf{J}}}_{\mathfrak{j},:,:}^{(0)}\ubar{\bm{\nu}}_t \right) + \z^U \right) + \right.\nonumber\\&\quad\quad\quad\quad\left. c_{t} + \bar{\mathbf{J}}_{\mathfrak{j},:,:}^{(0)}\bar{\bm{\omega}}_t + \ubar{\mathbf{J}}_{\mathfrak{j},:,:}^{(0)}\ubar{\bm{\omega}}_t \right), \label{eqn:global_bnd_worst_noise}
\end{align}
\endgroup
where \cref{eqn:global_bnd_sub_piUCL} substitutes the definition of $\piUCL$ from~\cref{thm:particular_xt}, \cref{eqn:global_bnd_observation} substitutes the observation from~\cref{eqn:closed_loop_dynamics}, \cref{eqn:global_bnd_linear_in_x} separates terms that depend on $\x_t$, and \cref{eqn:global_bnd_worst_noise} introduces the worst-case realizations of process and measurement noise.
Substituting $\Muj, \Nuj$ results in~\cref{eqn:global_upper_bnd_generic_optimization}.
The proof of the lower bound follows similarly.
\end{proof}

\end{lemma}


The optimization problems in~\cref{eqn:global_upper_bnd_generic_optimization,eqn:global_lower_bnd_generic_optimization} have convex cost with convex constraints $\xinX$ (e.g., polytope $\mathcal{X}_{in}$).
We solve the linear programs (LPs) with \texttt{cvxpy}~\cite{diamond2016cvxpy},
\begin{align}
    \gamma_{t+1,\mathfrak{j}}^U &= \texttt{LP}(\Muj\mathbf{x}_t, \Ain, \mathbf{b}^\text{in})+\Nuj \\
    \gamma_{t+1,\mathfrak{j}}^L &= \texttt{LP}(-\Mlj\mathbf{x}_t, \Ain, \mathbf{b}^\text{in})+\Nlj.
\end{align}

\subsection[Accounting for Control Limits]{Accounting for Control Limits, $\mathcal{U}_t$}\label{sec:approach:control_limits}

The key terms in \cref{thm:particular_xt} can be modified to account for control input constraints, as
\begin{align}
    \piUCL(\y_t) &= \textrm{Proj}_{\mathcal{U}_t}\left(\mathbf{\Upsilon}^{(0)}_{\mathfrak{j},:,:}\y_t+ \z^U\right) \label{eqn:piucl_control_constraints}\\
    \piLCL(\y_t) &= \textrm{Proj}_{\mathcal{U}_t}\left(\mathbf{\Xi}^{(0)}_{\mathfrak{j},:,:}\y_t+ \z^L \right) \label{eqn:pilcl_control_constraints},
\end{align}

A common example is box control input constraints.
The element-wise control input is,
\begin{align}
    \pi_{j,\mathfrak{j}}^{U_{CL}}(\y_t) =& \begin{cases}
        \mathrm{clip}(\pi_j^U(\y_t), \ubar{\mathbf{u}}_j, \bar{\mathbf{u}}_j), & \mathrm{if}\ \Aoutj  B_{t,:,j} \geq 0 \\
        \mathrm{clip}(\pi_j^L(\y_t), \ubar{\mathbf{u}}_j, \bar{\mathbf{u}}_j), & \mathrm{otherwise}
    \end{cases}
    \label{eqn:pi_upoly_elementwise_control_constraints},
\end{align}
where $\mathrm{clip}$ saturates the control if it exceeds the limits.
However, this could be non-convex depending on the domain of $\x_t$ (and violates DCP rules in~\texttt{cvxpy}~\cite{diamond2016cvxpy} regardless).
In this work, we only apply part of the control input constraint,
\begin{align}
    \pi_{j,\mathfrak{j}}^{U_{CL}}(\y_t) =& \begin{cases}
        \mathrm{min}(\pi_j^U(\y_t), \bar{\mathbf{u}}_j), & \mathrm{if}\ \Aoutj  B_{t,:,j} \geq 0 \\
        \mathrm{max}(\pi_j^L(\y_t), \ubar{\mathbf{u}}_j), & \mathrm{otherwise}
    \end{cases}
    \label{eqn:pi_upoly_elementwise_control_constraints_approx},
\end{align}
and raise an error if the other limit is violated (which did not happen in our experiments).
Future work will investigate solutions via convex relaxations \cite{yang2020graduated} of the $\mathrm{clip}$ function.

\subsection{Converting State Constraints into Reachable Sets}\label{sec:approach:reachable_sets}

\subsubsection[Reachable Sets as l-inf balls]{Reachable Sets as $\ell_{\infty}$ balls}

Assume $\mathcal{X}_0$ is an $\ell_p$ ball.
Define $\{p,\epsilon,\mathbf{x}_0\}$ s.t. $\mathcal{X}_0\subseteq\mathcal{B}_p(\mathbf{x}_0,\epsilon)$ and let $\bar{\mathcal{R}}_0(\mathcal{X}_0)=\mathcal{X}_0$.
Using the results of the previous section, use $\mathbf{x}_{t=0} \in\mathcal{B}_{p}(\mathbf{x}_0,\epsilon)$ to compute $(\gamma_{1,\mathsf{j}}^L,\, \gamma_{1,\mathsf{j}}^U)$ for each index of the state vector $\mathsf{j}\in[n_x]$, specifying $\Aout=\mathbf{I}_{n_x}$.
Recursively compute
\begin{align} \hspace*{-1.5mm}
    \bar{\mathcal{R}}_{t+1}(\mathcal{X}_0)=\mathcal{B}_{\infty}\left(\frac{\gamma_{t+1,:}^U + \gamma_{t+1,:}^L}{2}, \frac{\gamma_{t+1,:}^U - \gamma_{t+1,:}^L}{2}\right).
\end{align}

\subsubsection{Reachable Sets as Polytopes}

Assume $\mathcal{X}_0$ is an $\ell_p$ ball or polytope.
Either define $\{p,\epsilon,\mathbf{x}_0\}$ s.t. $\mathcal{X}_0\subseteq\mathcal{B}_p(\mathbf{x}_0,\epsilon)$ or define $\{\Ain, \mathbf{b}^\text{in}\}$ s.t. $\mathcal{X}_\text{in}=\{\mathbf{x}_t \lvert \Ainbin\}$ and let $\bar{\mathcal{R}}_0(\mathcal{X}_0)=\mathcal{X}_0$.
Using the results of the previous section, use $\mathbf{x}_{t=0} \in\mathcal{B}_{p}(\mathbf{x}_0,\epsilon)$ or 
$\{\Ain, \mathbf{b}^\text{in}\}$ to compute $(\gamma_{1,\mathfrak{j}}^L,\, \gamma_{1,\mathfrak{j}}^U)$ for each index of output polytope facets $\mathfrak{j}\in[m_{out}]$, giving
\begin{align}
    \bar{\mathcal{R}}_{t+1}(\mathcal{X}_0) = \{ \mathbf{x}_{t} \lvert \begin{bmatrix} \Aout \\ -\Aout \end{bmatrix} \mathbf{x}_t \leq \begin{bmatrix} \gamma_{t+1,:}^U \\ -\gamma_{t+1,:}^L \end{bmatrix} \}.
\end{align}
In both cases, $\bar{\mathcal{R}}_{t}(\mathcal{X}_0)\supseteq\mathcal{R}_{t}(\mathcal{X}_0) \forall t\geq0$, so these $\bar{\mathcal{R}}_{t}$ can be used to verify the original closed loop system~\cref{eqn:closed_loop_dynamics}.

\subsection{Tighter Reachable Sets via Partitioning the Input Set}

NN relaxation methods can be improved by partitioning the input set, particularly when the input set is large and of low dimension.
Here, we achieve tighter bounds by splitting $\mathcal{X}_0$ into several subsets, computing $N$-step reachable sets for each of the subsets separately, then returning the union of all reachable sets from each subset.
This idea falls into the general framework from~\cite{everett2020robustness} of choosing a propagator and a partitioner (e.g., uniform~\cite{xiang2018output}) for the analysis, where Reach-LP/SDP represent propagators for closed-loop systems.

%% file: results.tex

\section{Numerical Experiments}

\begin{figure}[t]
	\centering
	\begin{subfigure}{0.5\linewidth}
	    \centering
		\footnotesize{
		\begin{tabular}{|c|c|c|}
	         \hline
	         Algorithm & Runtime [s] & Error \\
	         \hline
	         Reach-SDP~\cite{hu2020reach} & 20.31 & 206 \\ 
	         Reach-SDP-Partition & 347.14 & \textbf{19.35} \\ 
	         Reach-LP & \textbf{0.63}& 848 \\ 
	         Reach-LP-Partition & 9.87 & 19.87 \\ 
	         \hline
	    \end{tabular}
	    }
	    \caption{Runtime vs. Error}
    \end{subfigure}\\
	\begin{subfigure}{0.5\linewidth}
		\centering
		\includegraphics[width=\linewidth]{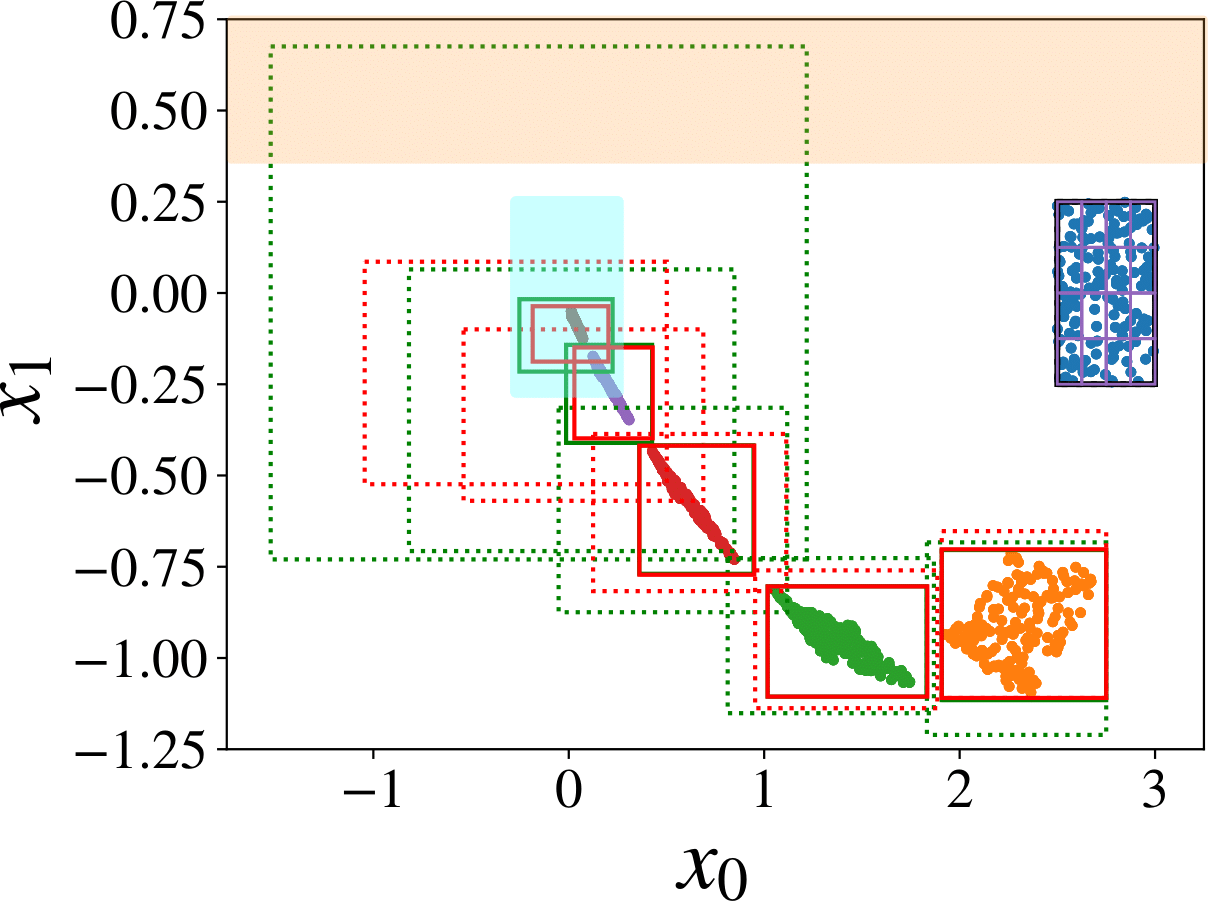}
		\captionsetup{justification=centering}
		\caption{Reachable Set Estimates}
		\label{fig:double_integrator_reachable_set_trajectory}
	\end{subfigure}%
	\begin{subfigure}{0.5\linewidth}
	    \includegraphics[width=0.95\linewidth]{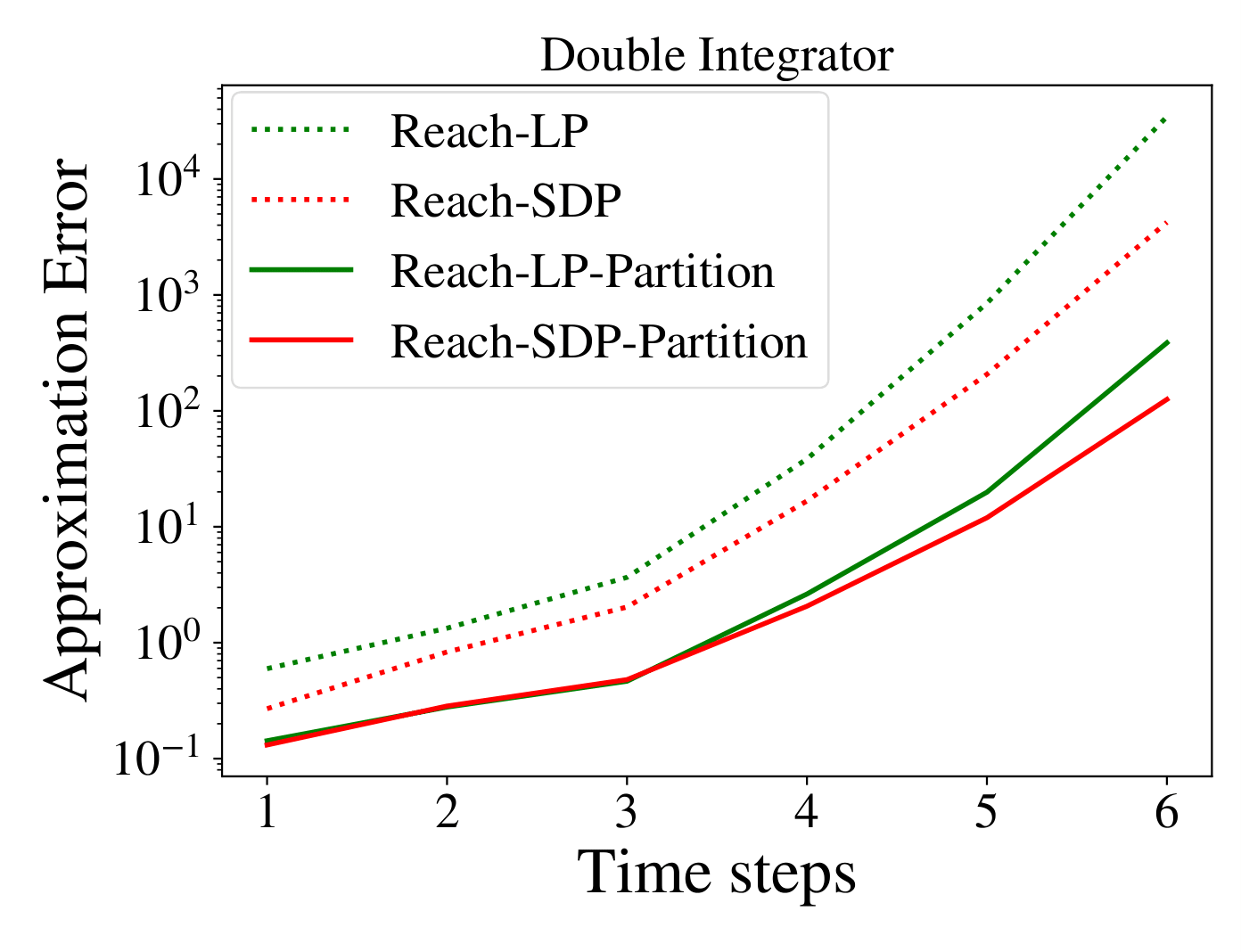}
	    \captionsetup{justification=centering}
	    \caption{Over-approximation error}
		\label{fig:double_integrator_reachable_set_error}
	\end{subfigure}%
	\caption{Reachable Sets for Double Integrator. In (a), Reach-LP is $30\times$ faster to compute but $4\times$ looser than Reach-SDP~\cite{hu2020reach}. Reach-LP-Partition refines the Reach-LP bounds by splitting the input set into 16 subsets, giving $10\times$ faster computation time and $2\times$ tighter bounds than Reach-SDP~\cite{hu2020reach}. In (b), all reachable set algorithms bound sampled states across the timesteps, starting from the blue $\mathcal{X}_0$, and the tightness of these bounds is quantified per timestep in (c).}
	\label{fig:double_integrator_reachable_set}
	\vspace{-0.2in}
\end{figure}

This section demonstrates our convex reachability analysis tool, Reach-LP, on simulated scenarios.
We first show an example verification task and quantify the improvement in runtime vs. bound tightness over the state-of-the-art~\cite{hu2020reach} for a double integrator system.
We then apply the algorithm on a 6D quadrotor model subject to multiple sources of noise.

\subsection{Double Integrator}\label{sec:results:double_integrator}

Consider the LTI double integrator system from~\cite{hu2020reach}, 
\begin{equation}
    \mathbf{x}_{t+1}=\underbrace{\begin{bmatrix}
        1 & 1 \\ 0 & 1
    \end{bmatrix}}_{A_t} \mathbf{x}_t + \underbrace{\begin{bmatrix}
        0.5 \\ 1
    \end{bmatrix}}_{B_t} \mathbf{u}_t,
\end{equation}
with $\mathbf{c}_t=0$, $C_t=I_2$ and no noise, discretized with sampling time $t_s=1s$.
As in~\cite{hu2020reach}, we implemented a linear MPC with prediction horizon $N_{MPC}=10$, weighting matrices $Q=I_2, R=1$, and terminal weighting matrix $P_{\infty}$ synthesized from the discrete-time Riccati equation, subject to state constraints $\mathcal{A}^C=[-5,5]\times[-1,1]$ and input constraint $\mathbf{u}_t\in[-1,1]\forall t$.
We used MPC to generate 2420 samples of state and input pairs then trained a NN with Keras~\cite{chollet2015keras} for 20 epochs with batch size 32.

\subsection{Comparison with Baseline}\label{sec:results:comparison_with_reachsdp}

\cref{fig:double_integrator_reachable_set} compares several algorithms on the double integrator system using a NN with [5,5] neurons and ReLU activations.
The key takeaway is that Reach-LP-Partition provides a $10\times$ improvement in reachable set tightness over the prior state-of-the-art, Reach-SDP~\cite{hu2020reach} (which does not use input set partitioning), while requiring $\frac{1}{2}$ of the computation time.
We implemented Reach-SDP in Python with \texttt{cvxpy} and \texttt{MOSEK}~\cite{andersen2000mosek}.
All computation times are reported from an i7-6700HQ CPU with 16GB RAM.

\cref{fig:double_integrator_reachable_set_trajectory} shows sampled trajectories, where each colored cluster of points represents sampled reachable states at a particular timestep (blue$\rightarrow$orange$\rightarrow$green, etc.).
Recall that sampling trajectories could miss possible reachable states, whereas these algorithms are guaranteed to over-approximate the reachable sets.
Reachable set bounds are visualized for various algorithms: Reach-SDP~\cite{hu2020reach}, Reach-LP, and those two algorithms after partitioning the input set into 16 cells.
The key takeaway is that while all approaches provide outer bounds on the sampled trajectories, the algorithms provide various degrees of tightness to the sampled points.

We quantify tightness as the ratio of areas between the smallest axis-aligned bounding box on the sampled points and the provided reachable set (minus 1), shown in \cref{fig:double_integrator_reachable_set_error} as the system progresses forward in time.
Note that as expected, all algorithms get worse as the number of timesteps increase, but that Reach-LP-Partition and Reach-SDP-Partition perform the best and similarly.
This provides numerical comparisons of the rectangle sizes from~\cref{fig:double_integrator_reachable_set_trajectory}.

Note that both Reach-LP and Reach-SDP methods could be improved by properly choosing the direction of polytope facets.
Additionally, while Reach-SDP can provide ellipsoidal bounds given the quadratic nature of the formulation, we implement only the polytope bounds in this comparison.

\subsection{Verification}\label{sec:results:verification}

A primary application of reachable sets is to verify reach-avoid properties.
In~\cref{fig:double_integrator_reachable_set_trajectory}, we consider a case with an avoid set $\mathcal{A}=\{\x\lvert \x_1\geq0.35\}$ (orange) and a goal set $\mathcal{G}=[-0.25,0.25]\times[-0.25,0.25]$ (cyan).
Each algorithm, except Reach-LP, verifies these properties for this 5-step scenario, highlighting the importance of tight reachable sets.

\subsection{Scalability to Deep NNs}\label{sec:results:scalability}
To demonstrate the scalability of the method, we trained NNs with 1-10 hidden layers of 5 neurons and report the average runtime of 5 trials of reachability analysis of the double integrator system.
In~\cref{fig:time_vs_num_layers}, while Reach-SDP appears to grow exponentially (taking $>800s$ for a 10-layer NN), our proposed Reach-LP methods remain very efficient ($<0.75s$ for Reach-LP on all NNs).
Note that we omit Reach-SDP-Partition ($\sim16\times$ more than Reach-SDP) from this plot to maintain reasonable scale.

\subsection[Ablation Study: Lp-balls vs. Polytopes]{Ablation Study: $\ell_{\infty}$ vs. Polytopes}\label{sec:results:ablation_lp_vs_polytope}

\begin{figure}
\centering
	\begin{subfigure}{0.54\linewidth}
	\centering
		\includegraphics[width=\linewidth, trim =20 20 40 20, clip]{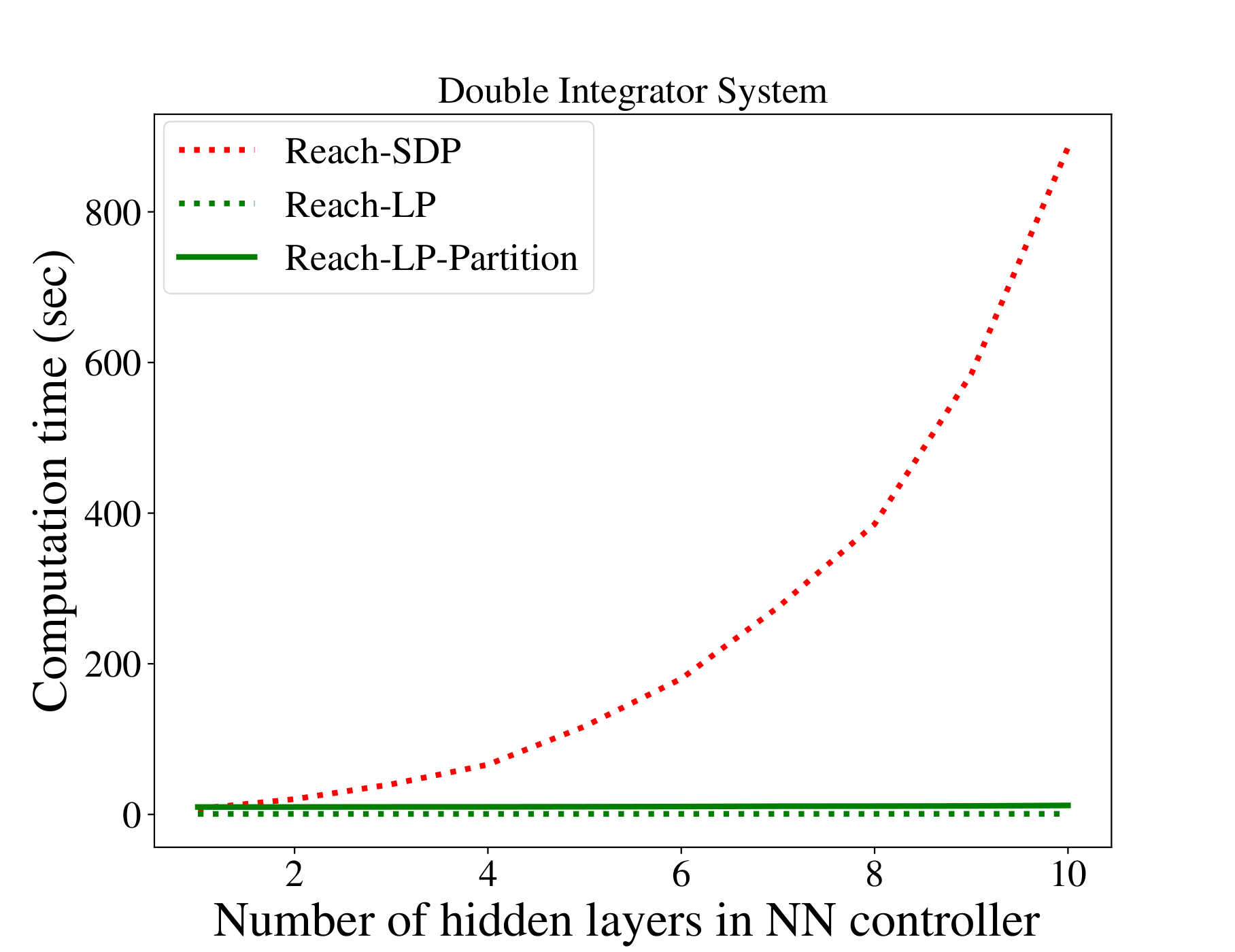}
		\caption{Runtime with NN depth}
	\label{fig:time_vs_num_layers}
	\end{subfigure}%
	\begin{subfigure}{0.46\linewidth}
	\centering
\includegraphics[width=\linewidth, trim =20 15 20 30, clip]{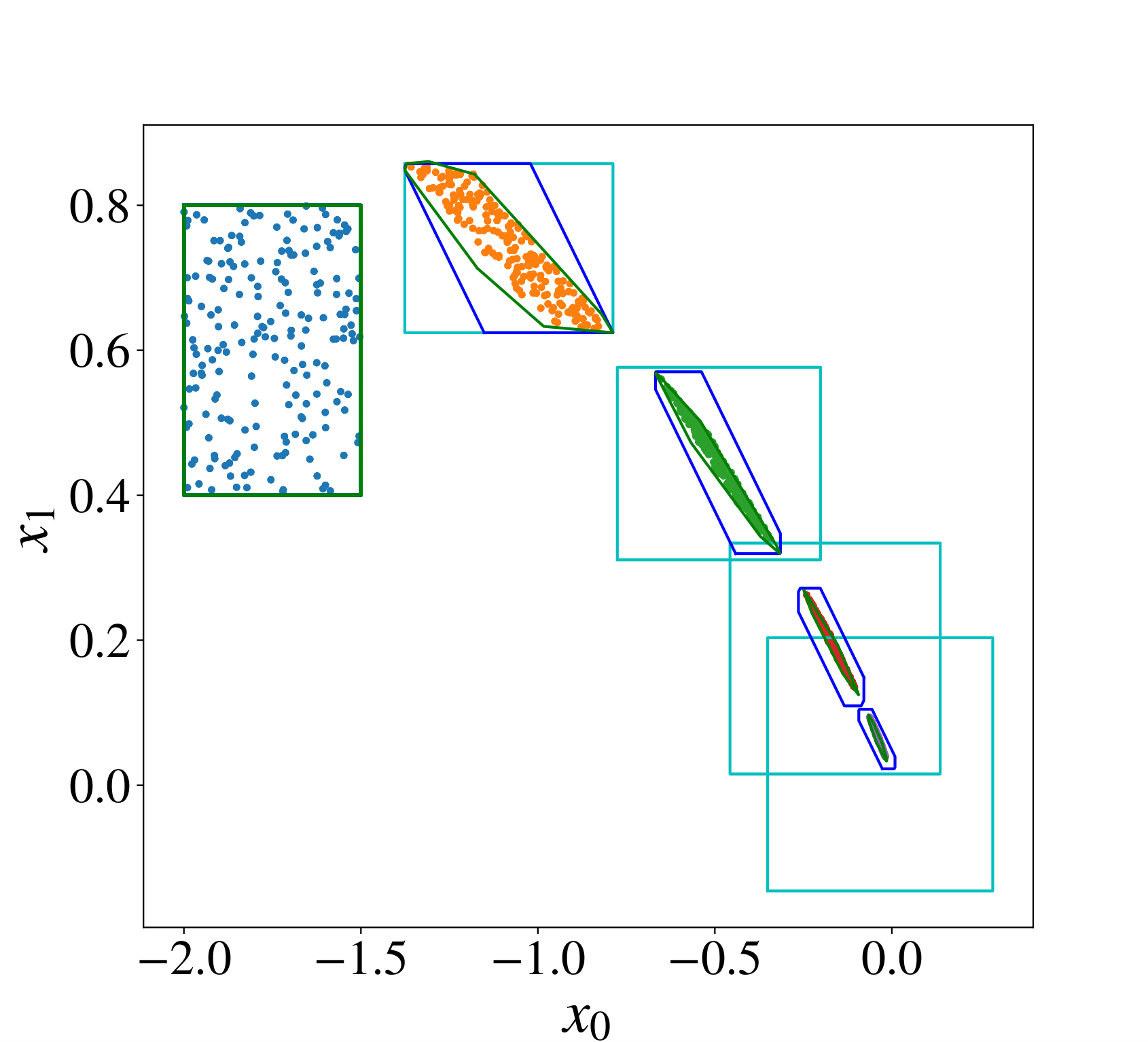}
    \caption{Number of Polytope Facets}
	\label{fig:double_integrator_reachable_set_polytope_facets}	
	\end{subfigure}
	\caption{(a) Our linear relaxation-based methods (Reach-LP, Reach-LP-Partition) scale well for deeper NNs (Reach-LP: 0.6 to 0.74s), whereas SDP-based methods grow to intractable runtimes. Note that input set partitioning multiplies computation time by a scalar.(b)  Using Reach-LP, the bounding shapes correspond to \textcolor{Cyan}{$l_{\infty}$-ball}, \textcolor{blue}{8-Polytope}, and \textcolor{ForestGreen}{35-Polytope}. Reachable sets become tighter with more facets.}
	\label{fig:double_integrator_result}
	\vspace{-0.2in}
\end{figure}
Recall that~\cref{sec:approach:reachable_sets} described reachable sets as either polytopes or $\ell_\infty$-balls.
\cref{fig:double_integrator_reachable_set_polytope_facets} shows the effect of that choice: as the number of sides of the polytope increases, the reachable set size decreases.
The tradeoff is that the computation time scales linearly with the number of sides on the polytope.
Note that a $\ell_\infty$-ball is a 4-polytope, and that $\mathcal{X}_0$ was chosen to show a different scenario than~\cref{fig:double_integrator_reachable_set}.

\subsection{6D Quadrotor with Noise}\label{sec:results:quadrotor}

\begin{figure}
	\centering
	\begin{subfigure}{0.4\linewidth}
		\centering
		\includegraphics[width=\textwidth, trim =80 0 100 0, clip]{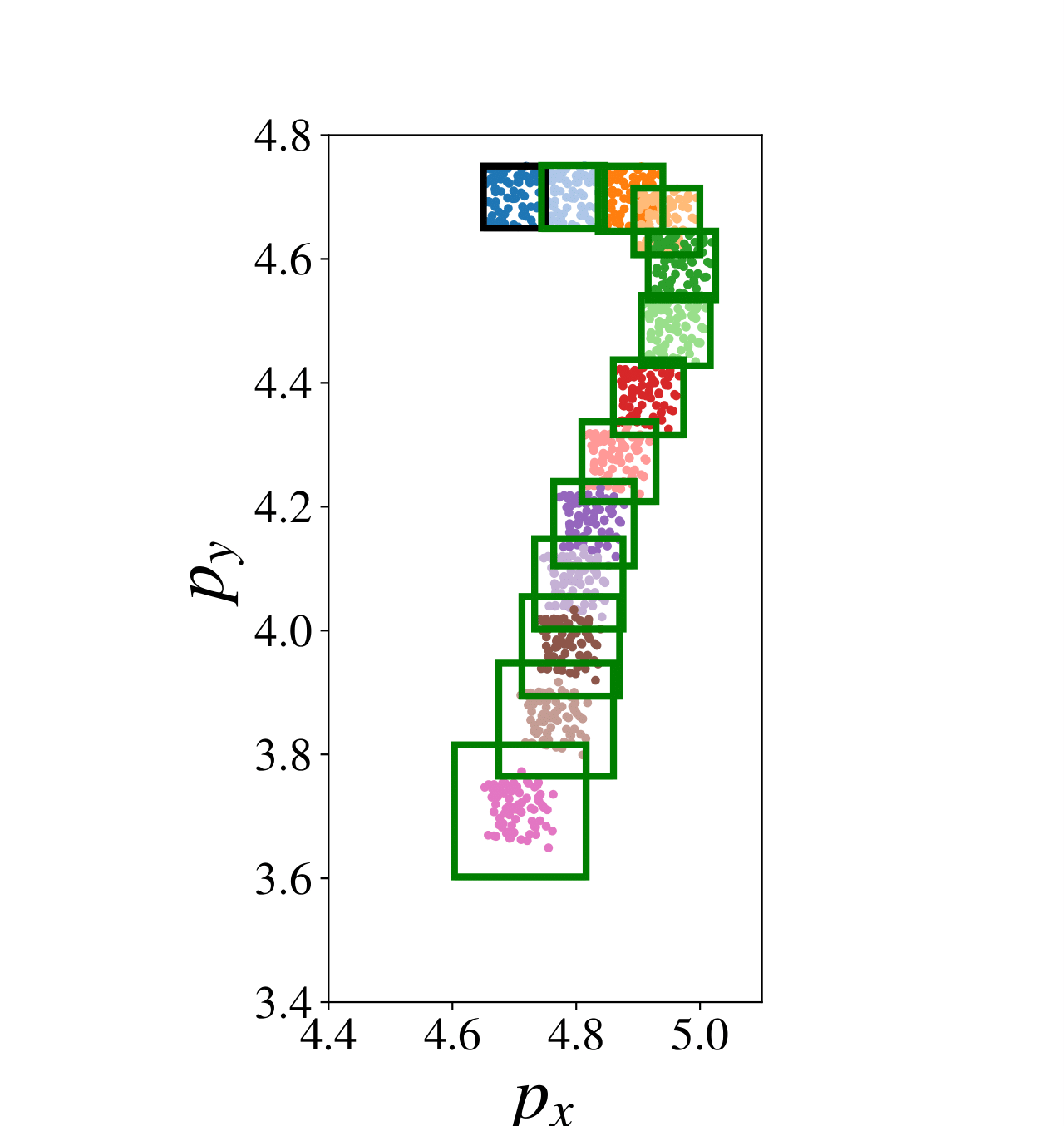}
		\captionsetup{justification=centering}
		\caption{No Noise}
		\label{fig:quadrotor_without_noise}
	\end{subfigure}%
	\hfill
	\begin{subfigure}{0.4\linewidth}
		\centering
		\includegraphics[width=\textwidth, trim =80 0 100 0, clip]{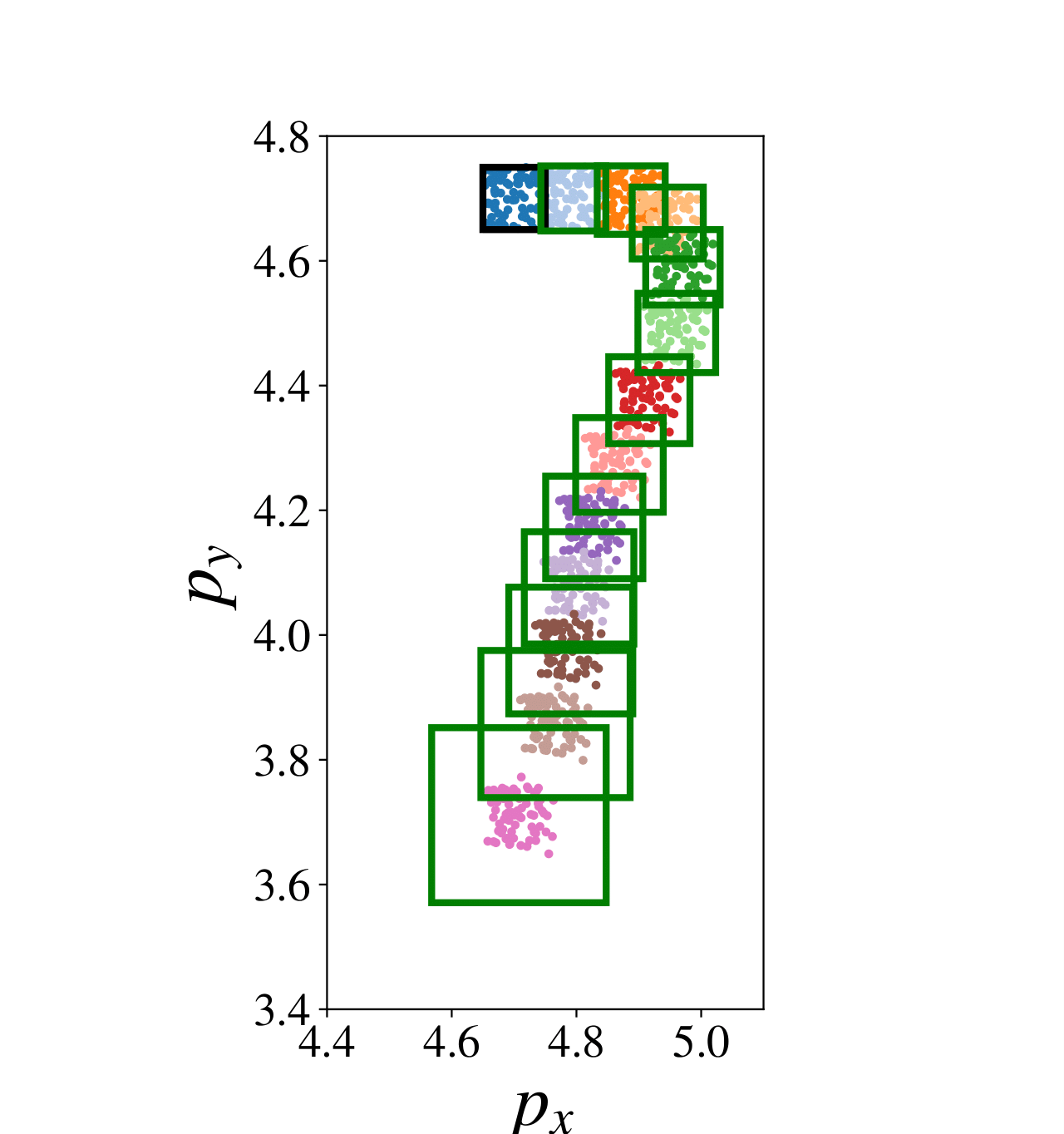}
		\captionsetup{justification=centering}
		\caption{Sensor \& Process Noise}
		\label{fig:quadrotor_with_noise}
	\end{subfigure}%
	\caption{Reachable Sets for 6D Quadrotor. Only $(x,y)$ states are shown, even though the reachable sets are computed in 6D. Green boxes (Reach-LP) bound the clusters of sampled points at each discrete timestep, starting from the blue $\mathcal{X}_0$. It took $4.89$ sec to compute the 12 reachable sets per scenario. In (b), $\bm{\nu}\sim\texttt{Unif}(\pm0.001\cdot\mathds{1}_6), \bm{\omega}\sim\texttt{Unif}(\pm0.005\cdot\mathds{1}_6)$.}
	\label{fig:quadrotor}
	\vspace{-0.2in}
\end{figure}

Consider the 6D nonlinear quadrotor from~\cite{hu2020reach,lopez2019verification}, 
\begin{align}
    \dot{\mathbf{x}}&=\underbrace{\begin{bmatrix}
        0_{3\times3} & I_3 \\ 0_{3\times3} & 0_{3\times3}
    \end{bmatrix}}_{A_t} \mathbf{x}_t + \underbrace{\begin{bmatrix}
         & g & 0 & 0 \\
        0_{3\times3} & 0 & -g & 0 \\
         & 0 & 0 & 1
    \end{bmatrix}^T}_{B_t} \underbrace{\begin{bmatrix}
        \text{tan}(\theta) \\
        \text{tan}(\phi) \\
        \tau
    \end{bmatrix}}_{\mathbf{u}_t} \nonumber\\ &\quad\quad+
    \underbrace{\begin{bmatrix}
        0_{5\times1} \\
        -g
    \end{bmatrix}}_{\mathbf{c}_t} + \bm{\omega}_t, \label{eqn:quadrotor_dynamics}
\end{align}
which differs from~\cite{hu2020reach,lopez2019verification} in that we add $\bm{\omega}_t$ as a uniform process noise, and that the output is measured as in~\cref{eqn:ltv_dynamics} with $C_t=I_6$, subject to uniform sensor noise.
As in~\cite{hu2020reach}, the state vector contains 3D positions and velocities, $[p_x,p_y,p_z,v_x,v_y,v_z]$, while nonlinearities from~\cite{lopez2019verification} are absorbed into the control as functions of $\theta$ (pitch), $\phi$ (roll), and $\tau$ (thrust) (subject to the same actuator constraints as~\cite{hu2020reach}).
We implemented a similar nonlinear MPC as~\cite{hu2020reach} in MATLAB to collect $(\mathbf{x}_t,\mathbf{u}_t)$ training pairs, then trained a [32,32] NN with Keras as above.
We use Euler integration to account for \cref{eqn:quadrotor_dynamics} in our discrete time formulation.

\cref{fig:quadrotor} shows the reachable sets with and without noise.
Note that while these plots only show $(x,y)$ position, the reachable sets are estimated in all 6D.
The first key takeaway is that the green boxes (Reach-LP with $\ell_\infty$-balls) provide meaningful bounds for a long horizon (12 steps, $1.2s$ shown).
Secondly, unlike Reach-SDP, Reach-LP is guaranteed to bound worst-case noise realizations.

%% file: conclusion.tex

\section{Future Directions}

Many open questions remain in analyzing closed-loop systems with NN controllers.
How to mitigate the conservatism due to the accumulation of approximation error over many timesteps?
Can similar methods be developed for nonlinear systems or systems with uncertainty in $A_t$ or $B_t$?
Can the ideas be extended naturally to continuous time systems, rather than through Euler integration?
How to handle the non-convex nature of saturations for control limits?
What partitioning scheme is best for closed-loop reachability?

\section{Conclusion}\label{sec:conclusion}

This paper proposed a convex relaxation-based algorithm for computing forward reachable sets of closed-loop systems with NN controllers.
Prior work is limited to shallow NNs and is computationally intensive, which limits applicability to real systems.
Furthermore, our method accounts for measurement of sensor and process noise as demonstrated on a quadrotor model.
The results show that this work advances the state-of-the-art in guaranteeing properties of systems that employ NNs in the feedback loop.